\documentclass[sigconf]{acmart}
\AtBeginDocument{%
  \providecommand\BibTeX{{%
    \normalfont B\kern-0.5em{\scshape i\kern-0.25em b}\kern-0.8em\TeX}}}

\usepackage{lipsum}
\usepackage{multirow}
\usepackage{soul}

\usepackage{float}
\usepackage{graphicx,lipsum,stfloats}

\usepackage[utf8]{inputenc}

\DeclareUnicodeCharacter{2318}{\applecmd}

\copyrightyear{2024}
\acmYear{2024}
\setcopyright{rightsretained}
\acmConference[UIST '24]{The 37th Annual ACM Symposium on User Interface Software and Technology}{October 13--16, 2024}{Pittsburgh, PA, USA} \acmBooktitle{The 37th Annual ACM Symposium on User Interface Software and Technology (UIST '24), October 13--16, 2024, Pittsburgh, PA, USA} \acmDOI{10.1145/3654777.3676378}
   
\acmISBN{979-8-4007-0628-8/24/10}

\begin{document}

\title[Vision-Based Hand Gesture Customization from a Single Demonstration]{Vision-Based Hand Gesture Customization from~a~Single~Demonstration}



\author{Soroush Shahi}
\affiliation{%
  \institution{Apple Inc.}
  \country{USA}
}
\email{shahi@apple.com}

\author{Vimal Mollyn}
\affiliation{%
  \institution{Apple Inc.}
  \country{USA}
}
\email{vimal_mollyn@apple.com}

\author{Cori Tymoszek Park}
\affiliation{%
  \institution{Apple Inc.}
  \country{USA}
}
\email{coripark@apple.com}

\author{Runchang Kang}
\affiliation{%
  \institution{Apple Inc.}
  \country{USA}
}
\email{runchang_kang@apple.com}

\author{Asaf Liberman}
\affiliation{%
  \institution{Apple Inc.}
  \country{Israel}
}
\email{asaf_liberman@apple.com}

\author{Oron Levy}
\affiliation{%
  \institution{Apple Inc.}
  \country{Israel}
}
\email{oron_levy@apple.com}

\author{Jun Gong}
\affiliation{%
  \institution{Apple Inc.}
  \country{USA}
}
\email{jun_gong@apple.com}

\author{Abdelkareem Bedri}
\affiliation{%
  \institution{Apple Inc.}
  \country{USA}
}
\email{bedri@apple.com}

\author{Gierad Laput}
\affiliation{%
  \institution{Apple Inc.}
  \country{USA}
}
\email{gierad@apple.com}

\renewcommand{\shortauthors}{Shahi et al.}
\begin{abstract}
  Hand gesture recognition is becoming a more prevalent mode of human-computer interaction, especially as cameras proliferate across everyday devices. Despite continued progress in this field, gesture customization is often underexplored. Customization is crucial since it enables users to define and demonstrate gestures that are more natural, memorable, and accessible. However, customization requires efficient usage of user-provided data. We introduce a method that enables users to easily design bespoke gestures with a monocular camera from one demonstration. We employ transformers and meta-learning techniques to address few-shot learning challenges. Unlike prior work, our method supports any combination of one-handed, two-handed, static, and dynamic gestures, including different viewpoints, and the ability to handle irrelevant hand movements. We implement three real-world applications using our customization method, conduct a user study, and achieve up to 94\% average recognition accuracy from one demonstration. Our work provides a viable path for vision-based gesture customization, laying the foundation for future advancements in this domain. 
\end{abstract}



\maketitle


\section{Introduction}
Hand gesture recognition for human-computer interaction has been studied for decades, with research and commercial applications spanning virtual and augmented reality, wearables, and smart environments~\cite{rautaray2015vision, chakraborty2018review, gillian2014gesture, gong2016wristwhirl, iravantchi2019beamband,
lien2016soli, wu2020back, aigner2012understanding}. 
Increasingly, the research community has had a strong desire to go beyond identifying a set of predefined gestures, and to enable users to customize and define their own~\cite{lou2017personalized, oh2013challenges, wobbrock2009user}. Customization brings a multitude of benefits, making gestures more memorable~\cite{nacenta2013memorability, dong2015elicitation}, effective ~\cite{ouyang2012bootstrapping, xia2022iteratively}, and accessible~\cite{
anthony2013analyzing, bilius2023understanding}. However, customization is difficult because it necessitates an efficient and user-friendly data collection step (colloquially referenced as a registration or demonstration step), while also confronting the long-standing challenge of learning from limited samples, known as few-shot learning (FSL) \cite{song2023comprehensive}. Specifically, FSL is difficult because models must synthesize prior knowledge with a small amount of new information without overfitting~\cite{parnami2022learning}. 

Various algorithms have been explored to tackle FSL challenges in gesture recognition~\cite{wan2013one, lu2019one, li2020one, xu2022enabling}. For example, transfer learning~\cite{zou2021transfer}, fine-tuning~\cite{cote2019deep}, and maximizing few-shots through data augmentation and generative techniques have been investigated~\cite{xu2022enabling}. However, the applicability of these methods is limited, especially when the originally trained gestures significantly diverge from the customized gestures. Additionally, different data types demand distinct augmentation strategies~\cite{wen2020time}. For example, augmentation methods applied to images may not be suitable for time-series sensor data~\cite{wen2020time}. Likewise, generative techniques have their own challenges, including hallucination errors~\cite{ji2023survey}, rendering them unreliable for data synthesis~\cite{mikolajczyk2018data}. In response, meta-learning~\cite{hospedales2021meta} has emerged as an alternative approach to address FSL challenges. Meta-learning is a training technique where a model can learn \textit{how to learn more effectively}~\cite{vanschoren2018meta}. Although meta-learning has been applied to various classification problems ~\cite{rahimian2021fs, zeng2021user}, we are the first to enable a vision-based gesture customization methodology from a single demonstration using this technique.


An equally important dimension of gesture customization is effective feature extraction, enabling users to freely create new and diverse gestures through limited demonstration. Notably, prior work in customization only supports a limited gesture set (\textit{e.g.,} only those with intense hand movements~\cite{xu2022enabling}), or relies on grammar-based visual declarative scripts~\cite{mo2021gesture}. Often overlooked is the capability to differentiate between irrelevant hand movements (\textit{i.e.}, background class) and intentional hand gestures. On the other hand, our approach allows users to perform gestures of various types, including one-handed, two-handed, static, and dynamic gestures, without needing to learn any complex grammar, and with the ability to exclude those hand movements that fall within the background class. We accomplish this by employing a feature extractor, notably a graph transformer model~\cite{zhou2022hypergraph}, which facilitates automatic feature extraction across diverse gesture types, regardless of the level of movement in the gesture (static vs. dynamic), number of hands involved (one- vs. two-handed), and camera viewpoints (egocentric~\cite{fathi2011understanding} vs. allocentric~\cite{klatzky1998allocentric}). To the best of our knowledge, we are the first to achieve this level of customization scope and flexibility, all from a single demonstration (see Section \ref{sec:Evaluation}). 

Initially, we pre-trained a graph transformer~\cite{zhou2022hypergraph} using a publicly available dataset~\cite{canavan2017hand} with 20 participants. Similar to previous work~\cite{de2016skeleton}, we use hand keypoints as an input feature to our model. Next, we employ a meta-learning algorithm coupled with meta-augmentation techniques to train a \textit{meta-learner}. We assess the utility of our gesture customization method through a user study and real-world applications. Initially, we gathered data from 21 participants (different from the 20 participants used in pre-training) who performed 20 distinct and varied gestures (see Figure~\ref{fig:AllGestures}). We leverage this dataset to assess the accuracy of our gesture customization method, achieving 94\%, 94\%, and 90\% average accuracy for two, three, and four new gestures (respectively), again from a single demonstration. 

Further, we also evaluate our method's adaptability to different viewpoints and a background class. Specifically, we employ a technique where we perform meta-learning using gestures captured from multiple viewpoints (\textit{i.e.,} egocentric and allocentric) and gestures captured from activities of daily living (\textit{e.g.}, washing hands), allowing it to reject the background class and learn a view-agnostic representation of a demonstrated custom gesture. Our results show that with this method, a custom gesture demonstrated once can be accurately distinguished from random hand movements and recognized regardless of viewpoint. We further highlight the results of this evaluation in Section~\ref{sec:Evaluation}.

Finally, we implemented three applications: (1) a tool for designers allowing them to experiment with custom gesture recognition models (see Figure~\ref{fig:DesignApp}), (2) a productivity add-on that augments keyboard shortcuts with custom gestures (see Figure~\ref{fig:deskviewdemo}), and (3) a mixed reality creativity application allowing users to interact using custom gestures (see Figure~\ref{fig:avpdemo}). Our findings demonstrate that our gesture customization method is highly usable and learnable, and empowers users to effortlessly incorporate new gestures that are personalized to their preferences and capabilities.

\begin{figure*}
    \centering
    \includegraphics[width=1\textwidth]{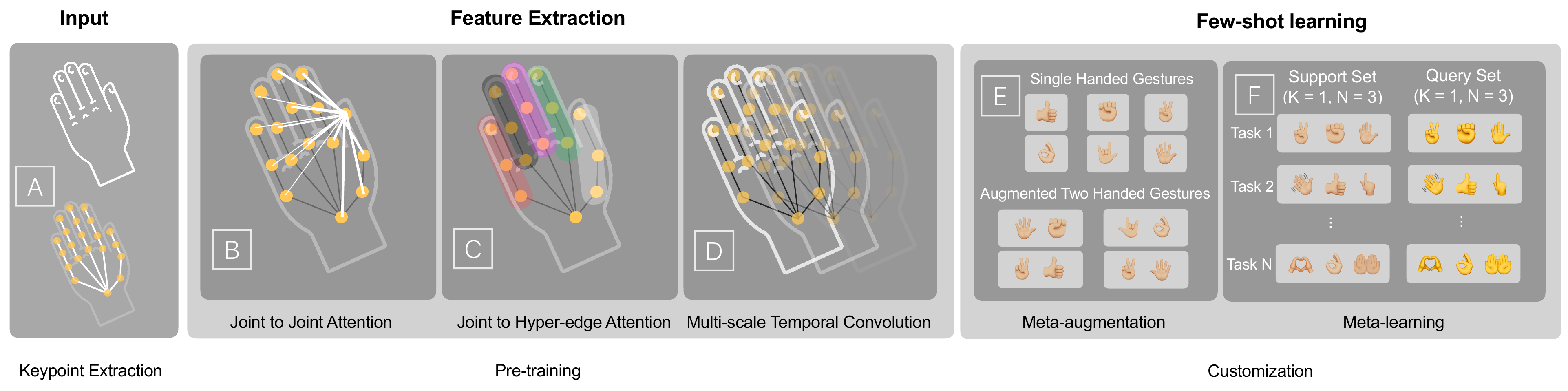}
    \caption{
    Overview of our method's pipeline: We use a hand pose estimation model (A) to extract hand skeleton and landmarks. We utilize a graph transformer to capture both (B) joint-to-joint and (C) joint-to-group attentions. Temporal features are extracted using (D) temporal convolutions. We implement and enhance customization and few-shot learning (FSL) through (E) meta-augmentation and (F) meta-learning approaches.
    }
    \label{fig:Pipeline}
    \Description{}
\end{figure*}
To summarize, our work makes the following contributions:
\begin{itemize}
    \item An end-to-end vision-based customization method that supports a variety of hand gestures such as static, dynamic, one-handed, and two-handed gestures, and multiple views such as egocentric and allocentric, using only one demonstration.
    \item A few-shot learning algorithm that combines meta-learning and meta-augmentation techniques with a graph transformer architecture. 
    \item An evaluation of our customization method with a background class, an ablation study, and comparisons with three baseline methods.
    \item Implementation of three applications demonstrating the practicality of our method in a real-world setting along with key insights relevant to the interactive experience of vision-based hand gesture customization. 
    
\end{itemize}

\section{Related Work}
\label{sec:RelatedWorks}
\subsection{Hand Gesture Recognition}

The ability to automatically recognize hand gestures has been a long-standing research problem in the human-computer interaction community. Over the years, researchers have explored various methods to detect hand gestures, including the use of different vision-based modalities (\textit{e.g.}, cameras, infrared ranging)~\cite{yasen2019systematic, kim2012digits}, acoustics~\cite{harrison2010skinput, iravantchi2019beamband}, and wearable inertial measurement units (IMUs)~\cite{akl2011novel, wen2016serendipity, laput2016viband}. Red-green-blue (RGB) cameras enable the detection of a wide variety of complex gestures and provide rich information to reliably detect hand gestures for many applications, such as sign language recognition~\cite{cheok2019review}. 

In this work, we focus on using the RGB camera modality to provide rich information supporting a diverse range of gestures. Researchers and practitioners have extensively explored static~\cite{kapitanov2022hagrid} and dynamic~\cite{canavan2017hand} hand gesture recognition using a single RGB camera~\cite{yang2019ddnet}. Among these works, many relied on raw RGB pixel values~\cite{wang2016large, lin2014human, mohanty2017deep} as input features, and some relied on hand skeleton (\textit{i.e.}, keypoints)~\cite{de2016skeleton}. Body skeleton features have been widely used for human action recognition as keypoints can be readily obtained through systems like the Kinect, LeapMotion, or robust vision-based pose estimation algorithms~\cite{doosti2019hand}. Skeleton-based approaches offer a more dependable alternative to traditional RGB- or depth-based methods, making these a promising choice for various real-world applications while also addressing certain computer vision challenges, like occlusion, illumination, and variations in appearance at a higher level. With the success of skeleton features in human action recognition and advancements in hand pose estimation, hand skeleton features have emerged as a promising asset for gesture recognition~\cite{de2016skeleton}.

Many prior works in gesture recognition have been tailored to specific types of gestures, like one-handed dynamic gestures, and typically assume access to extensive datasets of predefined gestures for classifier training~\cite{nunez2018convolutional, hou2018spatial}. However, there are many potential benefits of enabling user customization across a broader range of gestures, and with fewer user samples required. These advantages extend beyond increased memorability~\cite{nacenta2013memorability} and efficiency~\cite{ouyang2012bootstrapping} to include accessibility improvements for individuals with physical disabilities~\cite{anthony2013analyzing}. Customizing gestures necessitates the model to learn from just a few user-provided samples, which poses a unique challenge, particularly for deep learning models that demand substantial data volumes. Our work tackles these challenges by leveraging advanced transformer architectures~\cite{zhou2022hypergraph} and applying meta-learning algorithms~\cite{pmlr-v70-finn17a}.

\subsection{Gesture Customization}
Previous studies have developed tools that facilitate the generation of novel customized gestures (\textit{e.g.},~\cite{xu2022enabling, mo2021gesture}). The challenge with gesture customization is that it needs to support an efficient, yet effective, data collection process that creates an effortless experience for users. Although some methods require minimal training data to enable gesture recognition, they lack accuracy because they use traditional hand-crafted features~\cite{wu2012one, wan2015explore, konevcny2014one, escalante2017principal, zhang2017bomw}. Other works focus on supporting a specific type of gesture (\textit{e.g.}, one-handed dynamic)~\cite{xu2022enabling, stewart2020online}, or simplify gestures by translating the static component into hand shape and the dynamic component into palm motion~\cite{mo2021gesture}. However, for many gestures, this simplification results in a loss of important gestural information. For instance, a \emph{cutting scissors} gesture is dynamic, involving movement, but the palm is stationary and all movement occurs in the fingers. The \emph{pinch drag} gesture combines both palm and finger movements simultaneously, which must both be captured for accurate recognition. Our customization system addresses this complexity, accommodating a wide array of gestures, through an end-to-end computer vision pipeline that implicitly captures both spatial and temporal features using a transformer model.

We use public datasets to avoid hand-crafting features that suffer from low accuracy for complex hand gestures. Our method further supports customization by requiring only one demonstration in the registration process. Previous work explored one-shot gesture recognition using different vision modalities~\cite{lu2019one, li2021real}. Unlike prior work which used fine-tuning to learn new gestures, we use meta-learning techniques to overcome the challenges of learning from a few samples.

\subsection{Few-Shot Learning}
Few-shot learning is applicable in various human-computer interaction domains, notably to enhance user experience~\cite{dang2022prompt}, facilitate swift adaptation to user preferences~\cite{Park_2019_ICCV}, promote rapid prototyping and testing~\cite{mo2021gesture}, and effectively manage rare events or scenarios~\cite{zhao2022adaptive, lai-etal-2020-extensively}. Approaches to FSL can be categorized into two main groups: meta-learning-based~\cite{kulis2013metric, pmlr-v70-finn17a, sun2019meta} and non-meta-learning-based~\cite{torreytransfer, wang2019simpleshot, tian2020rethinking}, such as transfer learning and fine-tuning. In the context of gesture customization, Xu et al.~\cite{xu2022enabling} adopted a non-meta-learning approach, relying on transfer learning and fine-tuning with augmented data. However, naive fine-tuning can lead to overfitting~\cite{ying2019overview}, and augmenting skeleton data, depending on the type of gesture, may not always yield the desired results. For instance, rotation augmentation could potentially alter the class label, transforming a \emph{swipe right} into a \emph{swipe up} gesture. Previous research has applied FSL to gesture recognition in various domains, including electromyography~\cite{rahimian2021fs}, vision-based~\cite{wan2013one}, and WiFi-based methods~\cite{hu2021wigr}. In our study, we employ a model-agnostic meta-learning algorithm~\cite{pmlr-v70-finn17a}, initially designed for image classification tasks. We extend its capabilities by utilizing meta-augmentation techniques~\cite{rajendran2020meta} to generate new gesture classes. Subsequently, we fine-tune graph transformer meta-learners, resulting in better outcomes for gesture customization.

\section{Method}
An overview of our gesture customization pipeline is shown in Figure~\ref{fig:Pipeline}. First, we extract hand keypoints (Section~\ref{sec:keypoint_extraction}) and use a pre-trained graph transformer to produce hand gesture embeddings (Section~\ref{Sec:Pretraining}). To adapt to custom gestures, we then further train this model in a few-shot manner using meta-learning (Section~\ref{sec:Customization}). This allows the model to quickly adapt when a user registers a new set of gestures. We now describe these stages in detail. 
\begin{figure*}[t]
    \centering
    \includegraphics[width=1\textwidth]{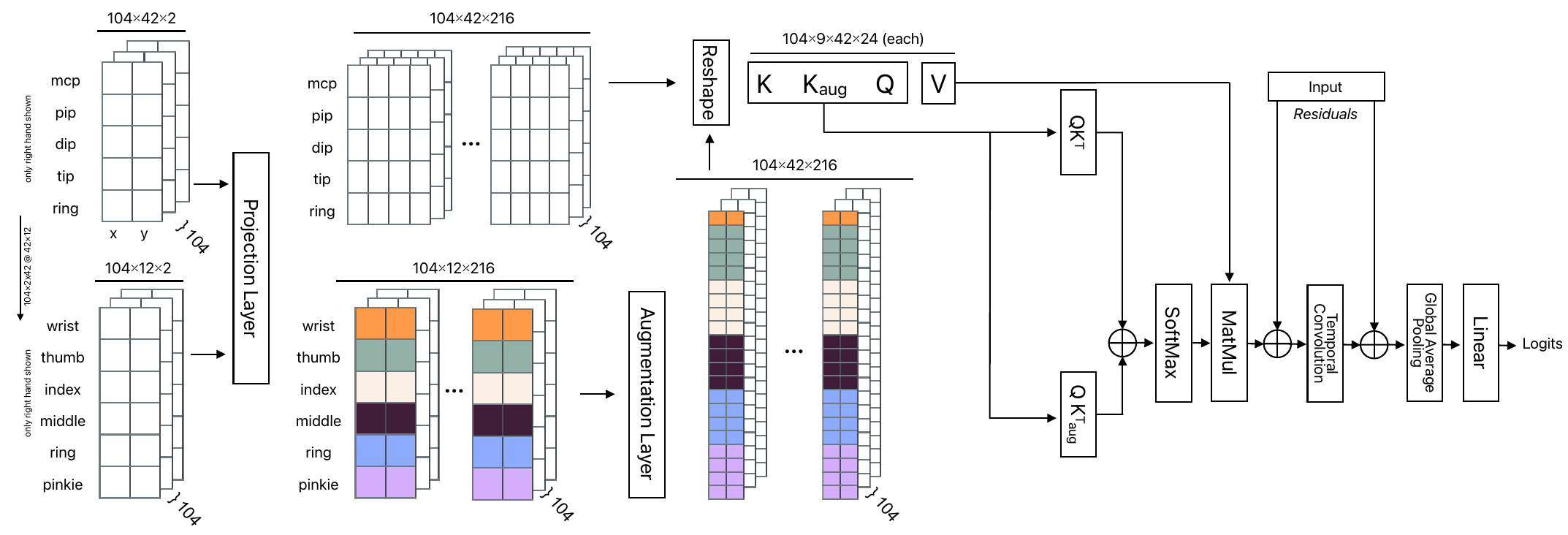}
    \caption{
    Graph transformer architecture used in our method. The graph transformer takes the keypoints and joint groups (fingers), projects them into higher dimensions and calculates attention maps to extract spatial features. The spatial and temporal features are then aggregated through temporal convolutions to yield the final output.
    }
    \label{fig:Architecture}
    \Description{}
\end{figure*}
\subsection{Keypoint Extraction}
\label{sec:keypoint_extraction}

Prior research has investigated various methods for representing hands, including the use of hand skeletons.
We employ a 2D hand pose estimation model~\cite{zhang2020mediapipe} to extract the hand's skeletal structure. The hand pose estimation model first detects and tracks the hand across a sequence of frames, and then extracts the keypoints for each hand detected in the frame. The skeleton comprises 21 key landmarks, including four points associated with each digit and one point for the wrist, with the location of each point specified using $x$- and $y$-coordinates. For two-handed gestures, keypoints are extracted for both hands resulting in 42 keypoints. We use these features as the input to our model.

\subsection{Pre-Trained Feature Extractor}
\label{Sec:Pretraining}
To lay the foundation for our gesture customization method, the initial step involves pre-training a model using a publicly accessible dataset. In this section, we describe the chosen public dataset along with the model’s architecture, and assess the performance of our pre-trained model when applied to a gesture classification task.

\subsubsection{Dataset}
The embedding space of the pre-trained model significantly influences the accuracy of the downstream task, such as customization. Consequently, it is vital to conduct pre-training using a high-quality dataset that offers diversity of classes and an abundance of samples. Additionally, collecting data from various individuals is essential to ensure that we capture variations between different subjects effectively. In recent years, several datasets have been made available to support researchers in enhancing vision-based gesture recognition techniques~\cite{kapitanov2022hagrid, materzynska2019jester, wan2016chalearn}. We have selected the dynamic hand gesture 14/28 dataset~\cite{Smedt_2016_CVPR_Workshops} for our work because it offers a rich assortment of 14 gesture categories characterized by diverse hand movements. Moreover, this dataset includes data from 20 individuals and provides two distinct methods for performing each gesture, such as using a single finger or the entire hand. 

We intentionally exclude five gestures (specifically, \emph{shake}, \emph{swipe left}, \emph{swipe right}, \emph{rotate clockwise}, and \emph{rotate counterclockwise}) from the training dataset because they were initially included for evaluating our customization method.

\subsubsection{Model Architecture}
As indicated in Section~\ref{sec:RelatedWorks}, prior studies have employed representations of the human body skeleton for action recognition. In our approach, we draw inspiration from this methodology, adapting a graph transformer model~\cite{zhou2022hypergraph} initially designed for human action recognition. Transformers have proven to be effective in tasks that rely on temporal features, and graph transformers, in particular, excel at representing skeletons (\textit{i.e.}, spatial information) by considering how each node (\textit{i.e.}, joint) relates to others using an attention mechanism (Figure~\ref{fig:Pipeline}). We have adjusted this model to make it work effectively for recognizing hand gestures. We create two separate hand graphs, one for the left hand and one for the right hand, and then concatenate them for the transformer model. In this graph, each node represents a specific landmark on each hand, and we establish connections between these nodes using the approach shown in Figure~\ref{fig:Pipeline}.
All the knuckles (MCPs) are linked to the wrist point. Each knuckle in a finger is connected to the corresponding finger's middle joint (PIP). Each middle joint (PIP) is connected to the end joint (DIP), and each end joint (DIP) is connected to the fingertip (TIP). Additionally, for every edge from some point $A$ to some point $B$, an edge from $B$ to $A$ is also added to the graph. Moreover, the graph transformer has the ability to compute an attention score for distinct groups of nodes. These node groups are defined based on the digits of the hand, creating separate groups for the thumb, index, middle, ring, and little fingers, allowing us to generate group-level attention maps.

To accommodate a broad range of gestures, we employ a top-down approach. The transformer model processes 104 frames (approximately 1.73 seconds with 60 frames per second, empirically determined), each containing 42 keypoints. For static gestures, we perform zero-padding on the temporal dimension, and for one-handed gestures, we perform "same"-padding on the spatial dimension (\textit{i.e.}, the keypoints of the present hand are duplicated to fill the place of the non-present hand). In contrast, for dynamic and two-handed gestures, actual data is populated in the temporal and spatial dimensions, respectively. This approach enables us to develop a single model capable of accommodating all gesture types. In the case of static gestures, which involve no motion, the system can expedite recognition by padding the input sequence with repeated frames to fulfill the 104 frames requirement. This method effectively reduces the delay for static gestures while maintaining the system's efficiency and accuracy.

\begin{figure*}[t!]
    \centering
    \includegraphics[width=1\textwidth]{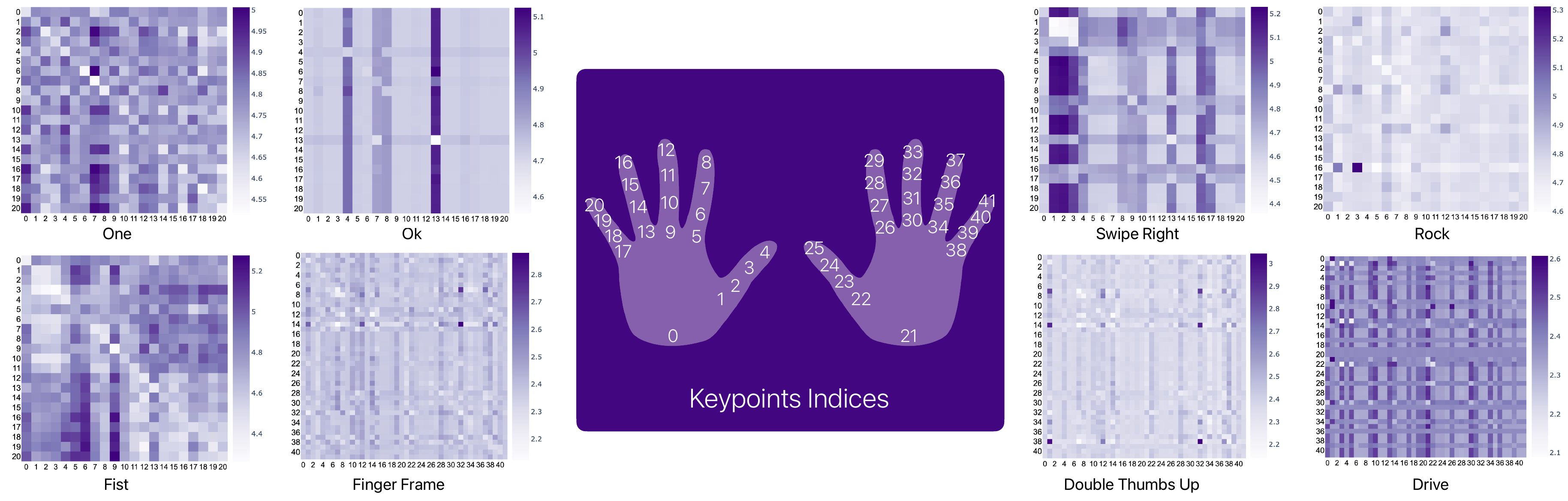}
    \caption{
    Visualization of the transformer's joint-to-joint attention maps for eight gestures. Each map is either a $21\times21$ (one-handed) or $42\times42$ (two-handed) matrix averaged across all attention heads ($N=9$). The model attends to different parts of the hand based on the gesture. For example, for a \emph{swipe right} gesture, most keypoints attend to the thumb and ring finger.
    }
    \label{fig:AttMaps}
    \Description{}
\end{figure*}

\subsubsection{Performance}
We use data from 15 participants for training and reserve 5 participants' data for validation. Our graph transformer model classifies the 9 gestures with an accuracy of 86\%. By employing joint-to-joint and joint-to-group attention, the graph transformer is able to learn spatial relationships between joints and fingers. This is in contrast to graph CNNs, and was validated in our experiments, in which we found the a graph CNN to be less accurate by a 6\% margin, similar to prior work~\cite{zhou2022hypergraph}.




Figure~\ref{fig:Architecture} highlights the process of computing these attention scores for both joint-to-joint and joint-to-group attention. In Figure~\ref{fig:AttMaps} we visualize the attention matrices by averaging the attention scores across all 9 attention heads. From these visualizations we can see that the model is able to learn unique spatial characteristics for each gesture. For example, in the case of the \emph{rock} gesture, one keypoint of the ring finger shows a high degree of attention towards one keypoint of the thumb. 


\subsection{Gesture Customization}
\label{sec:Customization}
In gesture customization, users provide a small set of their own gesture samples. Furthermore, the entire process, from registering the gesture to obtaining a trained model, should be fast, enabling users to perform multiple iterations as necessary. Deep learning models typically require substantial computation time, resources, and data. Attempting to fine-tune these models with only a handful of samples often results in less accurate outcomes. In this section, we describe the methods we employ to address these customization challenges, including limited data availability and compatibility with various types of gestures.

\subsubsection{Dataset}

\label{Sec:OurDataset}
As will be detailed in Section~\ref{sec:DataCollection}, our customization dataset comprises 20 distinct gesture classes including both static, dynamic, one-handed and two-handed gestures. Static gestures involve no hand movement, whereas dynamic gestures encompass various motions of the palm, fingers, or a combination of both. Each gesture class has ten demonstrations from each of the 21 participants resulting in a total of 4200 demonstrations. In addition to these 20 gesture classes, unlike prior work, we curated another dataset of "background" null classes, from videos of people doing everyday non-gestural activities like touching their hair, face, typing on a keyboard, washing the dishes etc. These gestures would serve as a way for our model to learn to reject the background class.
 \begin{figure*}[t]
    \centering
    \includegraphics[width=1\textwidth]{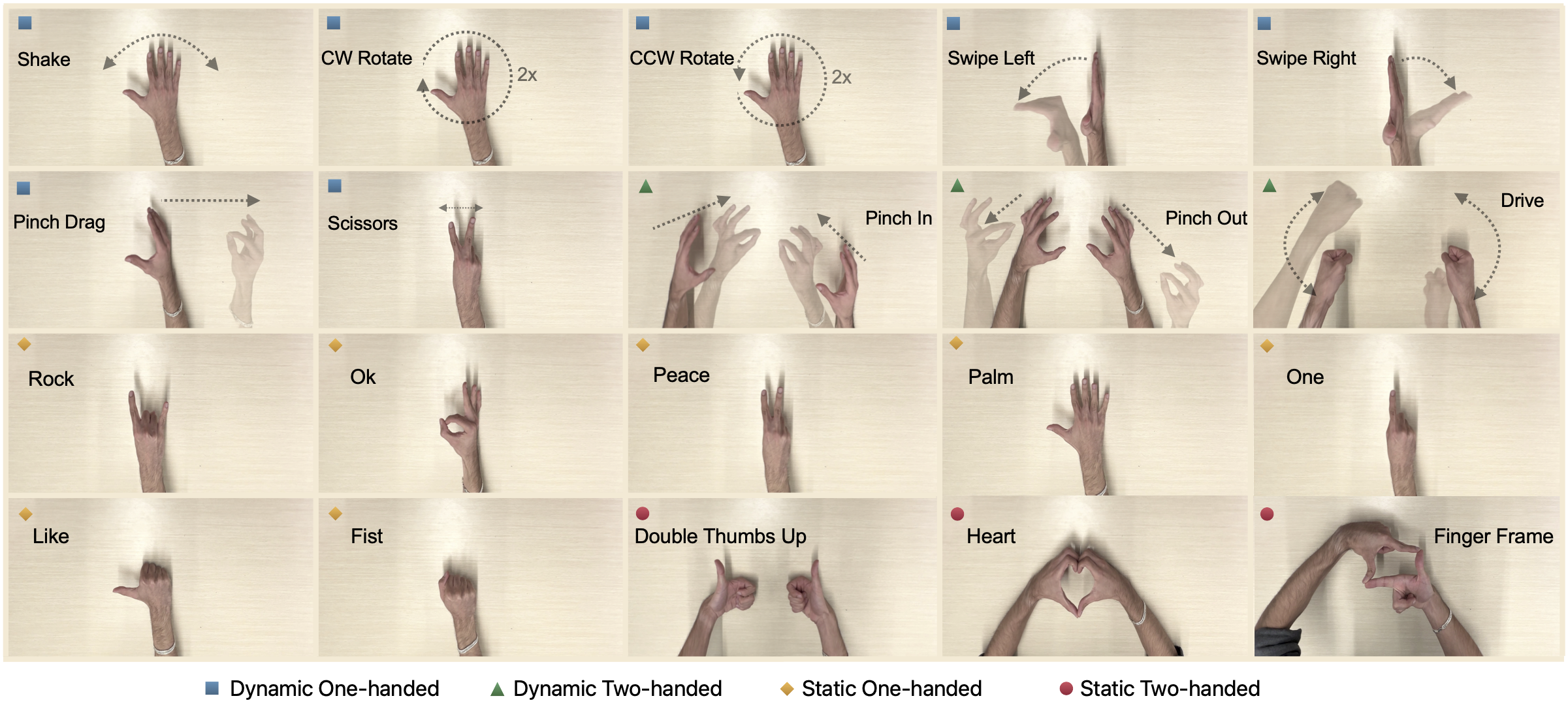}
    \caption{
    Shown here are the 20 gestures on which we evaluated our model. The blue square indicates the dynamic one-handed gestures, while the yellow diamond indicates the static one-handed gestures. Our dataset also comprises of six two-handed gestures, indicated by the green triangle for dynamic gestures and the red circle for static gestures. Our method can accommodate most gestures that a user might present from these gesture categories. This particular evaluation set was selected to represent the breadth of possible gestures, but should not be taken to suggest that our system can only accommodate these gestures.
    }
    \label{fig:AllGestures}
    \Description{}
\end{figure*}
\subsubsection{Meta-training}
In order to train our model to learn from very few demonstrations, we use meta-learning, which has emerged as a popular technique for few-shot learning. We use a model-agnostic meta-learning algorithm, specifically MAML~\cite{rajendran2020meta}, to continue training our graph transformer. At a high level, meta-learning enables a model to efficiently adapt to unseen classes with very few samples.


We use the customization dataset to train meta-learners. For a few-shot gesture classification problem involving $n$ custom gestures, the background class and $k$ demonstrations (\textit{i.e.}, ($n + 1$)-way, $k$-shot problem) for user $P$, we exclude $P$'s entire data from the training set along with the test split of the background class. We utilize the remaining $20-n$ classes from other 20 participants to generate new tasks and train the meta-learner. During each training iteration, we randomly choose $n$ classes and select $k$ samples at random from each of these $n$ classes. We also randomly choose $k$ samples of the background null class. These $k\times (n+1)$ samples form the support set, which the meta-learner uses to adapt the model to the specific task. Subsequently, we draw another set of $k \times (n+1)$ random samples  from the same classes (including the background class), forming the query set (non-overlapping with support set) for testing the model's performance, and calculate the test error. The model then takes a gradient step to minimize this error. Figure~\ref{fig:Pipeline}F illustrates the meta-learning process. The overall effect of this is that the model learns to quickly adapt to a new set of unseen gestures.

\subsubsection{Meta Augmentation}
\label{sec:metaaugmentation}
As with any deep-learning model, meta-learning is susceptible to overfitting, particularly when task diversity is limited. Augmenting is an effective method to tackle overfitting. However, classic augmentation methods, which are applied to individual data points (\textit{e.g.}, rotation), are not applicable in meta-learning as the model is overfitting on the tasks and not a specific set of classes. This concern becomes more pronounced when dealing with two-handed gestures, where our dataset contains only six distinct classes. To mitigate this challenge, we tackle the issue by generating two-handed gesture classes. These are created by combining two separate one-handed gestures, and are treated as novel tasks for training our meta-learner. During training time, we randomly replace the padded segment of a single handed gesture with another single handed gesture, creating a novel two-handed gesture class. Figure~\ref{fig:Pipeline}E shows the meta-augmentation process.

\subsubsection{Meta-Testing}
Once the model is trained on thousands of diverse tasks involving different gestures, it is now able to classify a new task involving new custom gestures. We sample $k$ demonstrations from each $n$ new custom gesture classes for user $P$ and adapt (\textit{i.e.}, fine-tune) the model to the new gestures. Then, we test the model on the remaining samples from the same user to calculate the accuracy. It is important to note that if a user defines a custom gesture that already existed in the training set, the recognition accuracy for that specific gesture is expected to improve, while the performance of other custom gestures remains unaffected.

\section{Evaluation}
\label{sec:Evaluation}
\subsection{Data Collection}
\label{sec:DataCollection}
We conducted a user study to collect data from 20 new gestures to train a customized model for each individual. A versatile gesture recognition system should accommodate a diverse range of gestures, each suited for specific functions. Static one-handed gestures are effective for initiating single events, whereas dynamic two-handed gestures provide a more intuitive means for continuous variable control. For example, a pinch and drag using two hands can be used to zoom in, or manipulate object size in a virtual reality application.

Following the approach outlined in the work by Choi et al.~\cite{CHOI2014171}, we referenced a taxonomy of static and dynamic hand gestures and selected a set of novel gestures that the pre-trained model had not encountered before. Figure~\ref{fig:AllGestures} illustrates these new gestures. Our selection criteria considered factors like hand and finger movements, as well as gesture similarities. For one-handed static gestures, we included widely recognized gestures such as \emph{thumbs up}, \emph{palm}, \emph{ok}, \emph{peace}, \emph{one}, \emph{rock}, and \emph{fist}, as they feature distinct finger configurations. In the realm of static two-handed gestures, we introduced \emph{double thumbs up}, akin to the \emph{thumbs up} gesture but performed with both hands. This choice allows us to evaluate our model's ability to distinguish between similar single and two-handed gestures. We also included \emph{heart} and \emph{finger frame}, two gestures involving distinct hand shapes. For dynamic gestures, we opted for gestures with similar shapes but varying hand movements, such as \emph{swipe left} and \emph{swipe right}, or \emph{rotate clockwise} and \emph{rotate counterclockwise}, as well as \emph{shake}. Additionally, we introduced \emph{scissors}, a gesture involving only finger movements, and \emph{pinch drag}, \emph{pinch in}, and \emph{pinch out}, which feature both hand motion and hand shape changes for either one or both hands. Some gestures, like \emph{driving} or \emph{rotate clockwise}, lack a predefined start and end state, as users may not consistently recall these details over time. We intentionally included these gestures to address this variability. 

\subsection{Participants and Apparatus}
\label{sec:ParticipantsAndApparatus}
As part of our evaluation, we conducted a data collection study with 21 participants (3 self-identified female, 18 male, with an average age of 34 $\pm$ 9). We employed two iPhone cameras to capture gestures from multiple angles, including egocentric and allocentric. To ensure synchronization and prevent any timing discrepancies, we connected these two iPhones to a MacBook Pro and recorded the gesture videos simultaneously at 60 frames per second.

\subsection{Design and Procedure}
The data collection process consisted of four sessions. The initial two sessions focused on static gestures. In these sessions, participants sequentially performed all 10 static gestures and then repeated each gesture nine more times in a random order. This randomness removes the temporal dependency between similar gestures enabling a more rigorous evaluation. A brief break was taken midway through these sessions. After concluding the static gestures data collection, participants took another break before proceeding to the two sessions dedicated to dynamic gestures. In the case of one-handed gestures, only one hand of the participant was visible in the scene, whereas for two-handed gestures, both hands of the participant were visible. The study took ~30 minutes for each participant to complete. Overall, 20 gestures were performed 10 times by 21 participants, resulting in 4200 demonstrations and approximately 500,000 frames.

\subsection{Background Class}
\label{sec:BackgroundClass}
In order to evaluate our model's ability to detect non-gesture background events, we also curated a dataset of background activities that users might perform day to day. We selected 600 video clips of people performing random actions like touching their hair, wiping their mouth, eating, typing on a keyboard, etc., and extracted hand keypoints for each of these clips. Similar to the method described in Section~\ref{sec:metaaugmentation}, we generated both 2 hand and single hand sequences to introduce further variance in our dataset. In total, we had 1400 sequences (150,000 frames) for training and a held out set of 360 sequences (30,000 frames) for testing. Due to the background class including hand movements that resemble gestures, gesture classification becomes particularly challenging. Nonetheless, Section~\ref{performance} reveals that our method's performance is not compromised by this additional class.

\begin{figure*}[t!]
    \centering
\includegraphics[width=1\textwidth]{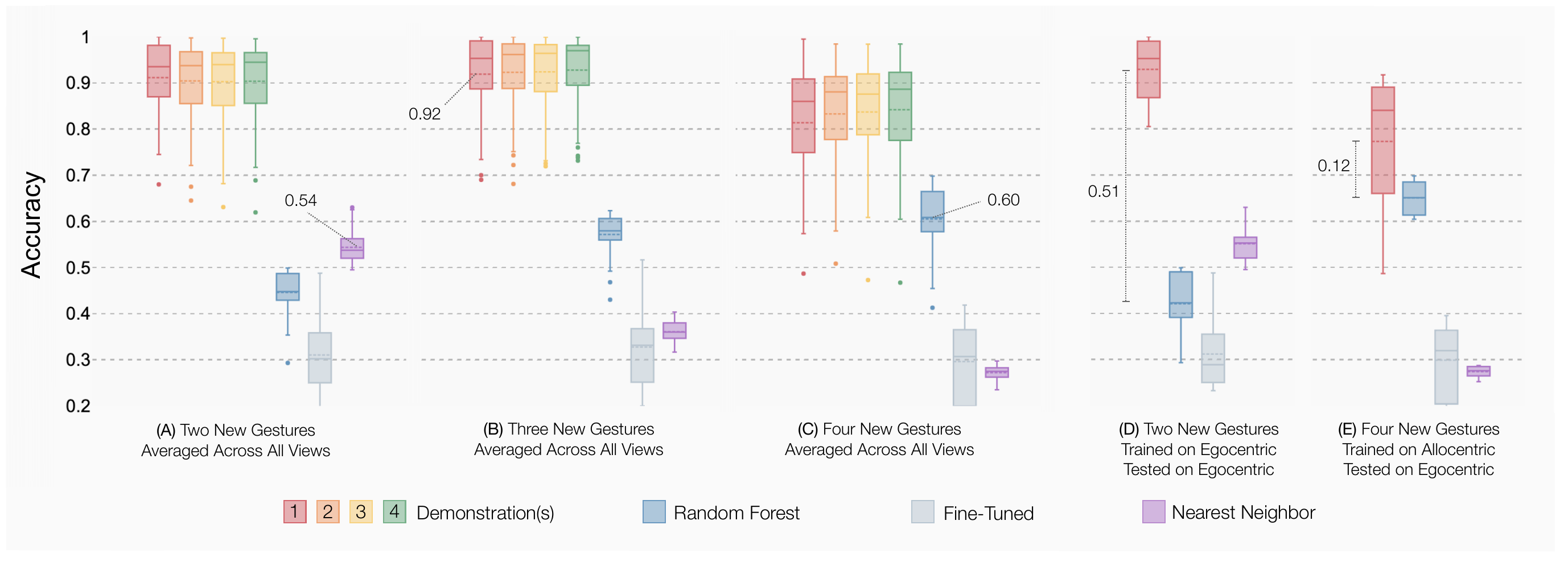}
    \caption{Accuracy of our method and baselines when tested with (A) two, (B) three, and (C) four new gestures along with the background class. In (D) we show a specific experiment where our method achieves its maximum accuracy improvement over baseline models such as the random forest. In (E) we show a specific experiment where our method achieves its minimum accuracy improvement over baseline models such as the random forest. For a complete evaluation and details, see Appendix~\ref{appendix:A}.}
    
    \label{fig:Results}
    \Description{}
\end{figure*}

\subsection{Training Details}
To train our model, we follow the meta-learning approach outlined in Section~\ref{sec:Customization}. We employ a stochastic gradient descent optimizer with a fixed learning rate of 0.025, and we continue training the meta-learners until they reach convergence. As mentioned before, each experiment is characterized as an \textit{n-way k-shot} problem, where \emph{n} denotes the number of new gestures, and \emph{k} signifies the number of demonstrations provided in the support set. To challenge the generalizability of our method, we employ the leave-one-subject-out cross validation approach. We evaluate our approach across a range of scenarios, including $n=2$, $n=3$, $n=4$, and $0<k<5$. To test all possible combinations of gestures would necessitate an impractical and computationally infeasible number of experiments, estimated at ${20 \choose 2}\cdot{20 \choose 3}\cdot{20 \choose 4}\cdot21$. To address this challenge, we take a hybrid approach where we manually include certain gestures and fill in the empty spots with random gestures. In curating these experiments, we consider factors such as gesture similarity in each combination. Initially, we group together similar gestures that pose a greater challenge for differentiation. For instance, gestures like \emph{pinch in} and \emph{pinch out}, or \emph{swipe left} and \emph{swipe right}, are paired, ensuring they appear together within the customized gesture set. When customization extends to more than two classes (\textit{e.g.}, 3 or 4), we randomly select an additional gesture(s) to complement the user's custom gesture set. The rest of the gesture classes are used to train the meta-learner. In total we curate 30 experiments covering all the 20 gestures, and all the 21 participants in our dataset.

\subsubsection{Training with a Background Class}
Most prior work assume gesture candidates to be segmented before classification, leading to difference in experimental and real-world performance. Instead, we additionally train our model on classifying the $n$ selected gestures plus a background class, which we believe more closely resembles real-world performance.

\subsubsection{Training with Multiple Views}
Our dataset includes gestures captured from egocentric and allocentric views. We leverage that during training, and randomly sample gestures from various viewpoints, enabling the model to learn a view-agnostic representation of the gestures. As a result, when it comes to the testing phase, the model is able to identify user's custom gestures from multiple views. To assess this capability, we train the model with a single demonstration from one viewpoint (\textit{e.g.}, egocentric) and subsequently evaluate its performance using data collected from another viewpoint (\textit{e.g.}, allocentric). 



\subsection{Model Performance}
\label{performance}
In this section, we present a summary of our customization method's accuracy when dealing with two, three, and four new gestures and when tested with gestures captured from same view and with different views alongside the baselines. Unlike previous work, we evaluate our models with an extra background class and show that our meta-learning method outperforms baselines by a significant margin. Figure~\ref{fig:Results} A-C shows a condensed summary of our evaluation results. For a complete evaluation, see Appendix \ref{appendix:A}.

\subsubsection{Using Same View Point}
 When the model is trained and tested with a single view (\textit{i.e.}, egocentric, allocentric) and, with just a single demonstration, our method attains 94\% accuracy on average in classifying two and three new gestures. When confronted with four new gestures, our method maintains 90\% accuracy. Furthermore, when we employ four demonstrations for training, the method demonstrates even higher performance, achieving average accuracy of 95\% for four new gestures. An example of this improvement is plotted in Figure~\ref{fig:Results} A-C.

\subsubsection{Using Different View Points}
When the model is trained on a particular view and tested on a novel view, our method attains 91\% accuracy on average in classifying two new gestures. Similarly, for three new gestures, our accuracy remains at 90\%. When confronted with four new gestures, our method maintains 82\% accuracy. Furthermore, when we employ four demonstrations for training, the method demonstrates even higher performance, achieving average accuracy rates of 95\%, 96\% and 93\% for two, three and four new gestures, respectively.

\subsubsection{Baselines}
We demonstrate the significance of meta-learning in facilitating gesture customization through an ablation study, in which we compare our method against three baseline approaches. First, we \emph{fine-tune} the model with just a single demonstration of the custom gesture and subsequently evaluate its performance on test samples. Second, we employ the nearest neighbor method within the feature embedding space (size=216) of the pre-trained model. This method calculates the pair-wise distance between the embeddings of the training and test gestures and matches the closest ones to compute accuracy. Third, we employ a traditional machine learning method and measure the performance of a random forest model in classifying gestures. The model which is fine-tuned on one demonstration achieves a maximum of 64\% accuracy in classifying two new gestures. Similarly, using the nearest neighbor method in the feature embedding space, the model achieves a maximum of 57\% accuracy in classifying two new gestures. Thus, our customization method using meta-learning improves over these deep-learning based baselines by \emph{at least} 30\%. Similarly, our method surpasses the random forest model by an average accuracy margin of 6\% across different numbers of gestures. The effectiveness of the random forest model can be attributed to its design, which is more resistant to overfitting compared to deep learning approaches. However, we show that when a non-gestural background class is introduced, adding complexity to the classification task, our method outperforms this baseline by an average of 34\%.

\subsubsection{Evaluation with a Background Class}
As discussed in Section~\ref{sec:BackgroundClass}, accounting for random and non-gestural hand movements which are not included in the registered gesture set is crucial for evaluating the model's effectiveness. Our results demonstrate that our method maintains robustness against the background class, losing a maximum of 5.5\% accuracy when confronted with four new gestures. This contrast is more pronounced with other baseline methods, underscoring the superiority of our meta-learning approach. The accuracy reduction when using the baseline methods is more significant, for example, the accuracy of the random forest model is decreased 39\% on average across two, three, and four gestures. This is depicted in Figure~\ref{fig:Results}D-E.

\subsubsection{Registration Time vs. Accuracy}  We analyzed the balance between demonstration count and average registration time, exclusive of user interface interaction. Our findings show that the average time for registering a single gesture is 1.88 $\pm$ 0.21 seconds. Thus, we estimate that registering a gesture with two, three, and four demonstrations takes about 3.8, 5.6, and 7.5 seconds, respectively, for a mix of static and dynamic types. Our results show that for two new gestures, more demonstrations do not improve accuracy significantly, but as the number of new gestures increases, more demonstrations do improve the accuracy of the model. For example, Figure \ref{fig:Results}A-C illustrates the method's accuracy with varying demonstration counts.

\subsubsection{Model's latency}
We measured the training and inference speed of our model on an Apple M1 Pro CPU. The unquantized version of our method takes, on average, 20 seconds to train a model for one custom gesture and about one minute to train for four gestures, with each training session running for five epochs. The average inference latency is 160 milliseconds on the same machine. Our implementation uses a 1.73-second sliding window with a step size of 100 milliseconds, resulting in a total inference time of approximately 1.89 seconds per window. These parameters can be optimized to decrease training and inference times by condensing epoch lengths, adjusting window sizes, and utilizing GPUs.

\section{Example Applications}
\subsection{Design Tool}

\begin{figure}[b]
    \centering
    \includegraphics[width=0.99\linewidth]{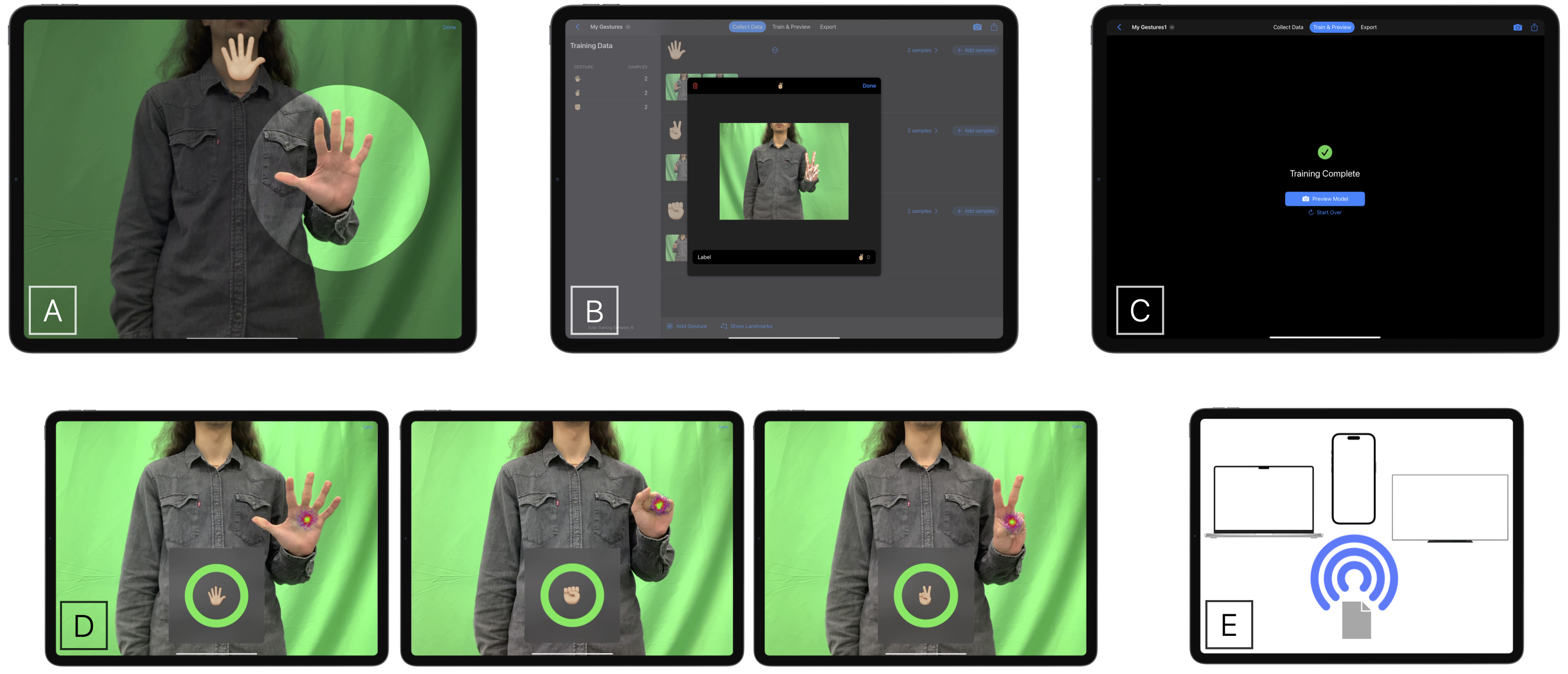}
    \caption{
    The overview of our gesture design tool. (A) The designer starts with registering their own custom gestures by demonstrating each gesture once. (B) The designer can review the samples collected, visualize the landmarks, and decide to delete, keep or add samples. (C) A graph transformer model is trained on-device using the designer's collected samples in under a minute. (D) The designer can preview and assess the real-time performance of the model. Based on the result, the designer can decide to edit the gestures or samples by going back to step AB. (E) The custom gesture model can be further be exported to other devices or applications enabling seamless interaction and deployment.
    }
    \label{fig:DesignApp}
    \Description{}
\end{figure}

Despite extensive research on hand gesture recognition and interaction robustness, there has been minimal emphasis on developing custom gesture design tools that require minimal data. Consequently, designers often restrict themselves to simple gestures, hindering the creation of applications, such as in game design or 3D drawing~\cite{lv2015touch, mazouzi2016ghost, dudley2018bare}, that benefit from more complex hand gestures. 

This section outlines the development of a gesture design tool aimed at facilitating the quick prototyping of hand gestures through our customization approach (see Figure~\ref{fig:DesignApp}).  We developed an iPad application using Swift UI and other iOS frameworks such as CoreML, featuring several key functionalities.
\subsubsection{Registration} This feature enables designers to define custom gestures with a single demonstration. The app achieves rapid and seamless registration by tracking hand movements and capturing motion-based samples. We designated a specific area of the screen to detect when a hand enters this zone. The app then uses the hand's motion within this area to identify the beginning and conclusion of a gesture, as illustrated in Figure~\ref{fig:DesignApp}A.

\subsubsection{Review} Designers have the opportunity to examine and modify the gestures they have registered. During this phase, they can add, delete, or alter the gesture labels and the associated samples. This step is demonstrated in Figure~\ref{fig:DesignApp}B.
\subsubsection{Training and Preview} Using the data they have collected, designers can train a model in under a minute. Then, a live preview allows designers to assess the trained model's performance in real-time (see Figure~\ref{fig:DesignApp}C-D).
\subsubsection{Export and Deployment} Unlike any other design tools, our application supports exporting the trained model in a CoreML format, making it compatible with any application that accepts this format (see Figure~\ref{fig:DesignApp}E). This feature was used by designers to incorporate custom gestures into other applications.

We provided the tool to a group of designers at Apple Inc. who are interested in quick hand gesture prototyping for evaluation purposes. Over six months, these designers utilized the tool to develop and test custom interactive gestures. They were particularly pleased with the simplicity of creating a custom model and its integration into various applications. The possibility of enhancing the tool was also discussed. An interest in expanding the tool's capability to include body pose customization was expressed, which is feasible with minor adjustments. For instance, hand landmarks might be substituted with body landmarks to accommodate this enhancement. Currently, the tool necessitates the user's touch interaction to alternate between gestures during the registration phase. The designers recommended a more natural transition between gestures using voice commands like "next."
\begin{figure}[b]
    \centering
    \includegraphics[width=\linewidth]{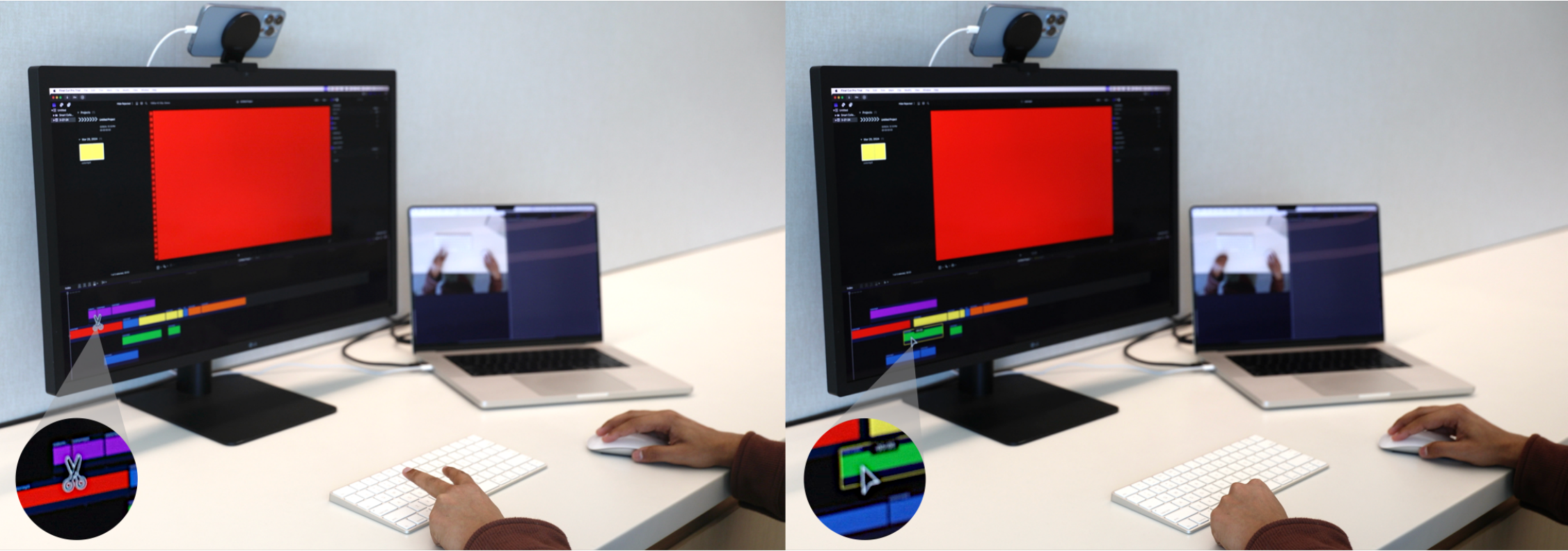}
    \caption{(Left) An example video editing application where the user has customized a left-hand gesture to invoke the blade tool. (Right) The system also recognizes null classes and changes the tool back to the default pointer.}
    \label{fig:deskviewdemo}
    \Description{}
\end{figure}

\subsection{Video Editing}
Creative tools, such as video editing software, often contain numerous toolbars and menus filled with a plethora of tools which can be daunting and inefficient for users to switch between. For this reason, creative professionals often memorize various non-intuitive keyboard shortcuts to quickly toggle between these tools. We envisioned a system in which the tools could be mapped to custom gestures that the user found intuitive. In this demo application, we utilize a smartphone mounted on a monitor and used Apple's Desk View technology to track the hands on the desk. We also modified a video editor to additionally take in hand gestures as input (see \autoref{fig:deskviewdemo}). The user can define an intuitive custom gesture, like the peace sign, and associate it with the action of splitting a video clip which is commonly accessible with ctrl/cmd+k shortcut. Similarly, the user can customize new gestures to select other tools, thereby reducing the burden of learning and memorizing a large number of awkward keyboard shortcuts. Additionally, when a null gesture (\textit{i.e.}, fist) is recognized, the tool changes back to a default pointer.

\subsection{Mixed Reality}

Using hand gestures as a primary mode of input has become common with the prevalence of mixed reality or spatial computing devices, such as the Apple Vision Pro, Meta Quest 3, and others. While these devices can free the user from hardware such as a mouse, keyboard, or controller, that freedom can also bring a lack of precision during interactions. When using a creative application like a photo editor, many tools are available, but they are often concealed within menus and require keyboard shortcuts to efficiently access. To further complicate the issue, users may have different subsets of tools that they use most frequently. Gesture customization can be of great use in this scenario, allowing users to specify custom gestures for the tools they reach for most often. This takes advantage of hand gestures as the natural input mechanism of these mixed reality devices and avoids the difficulty of navigating complex menus.

\begin{figure}[t]
    \centering
    \includegraphics[width=0.48\linewidth]{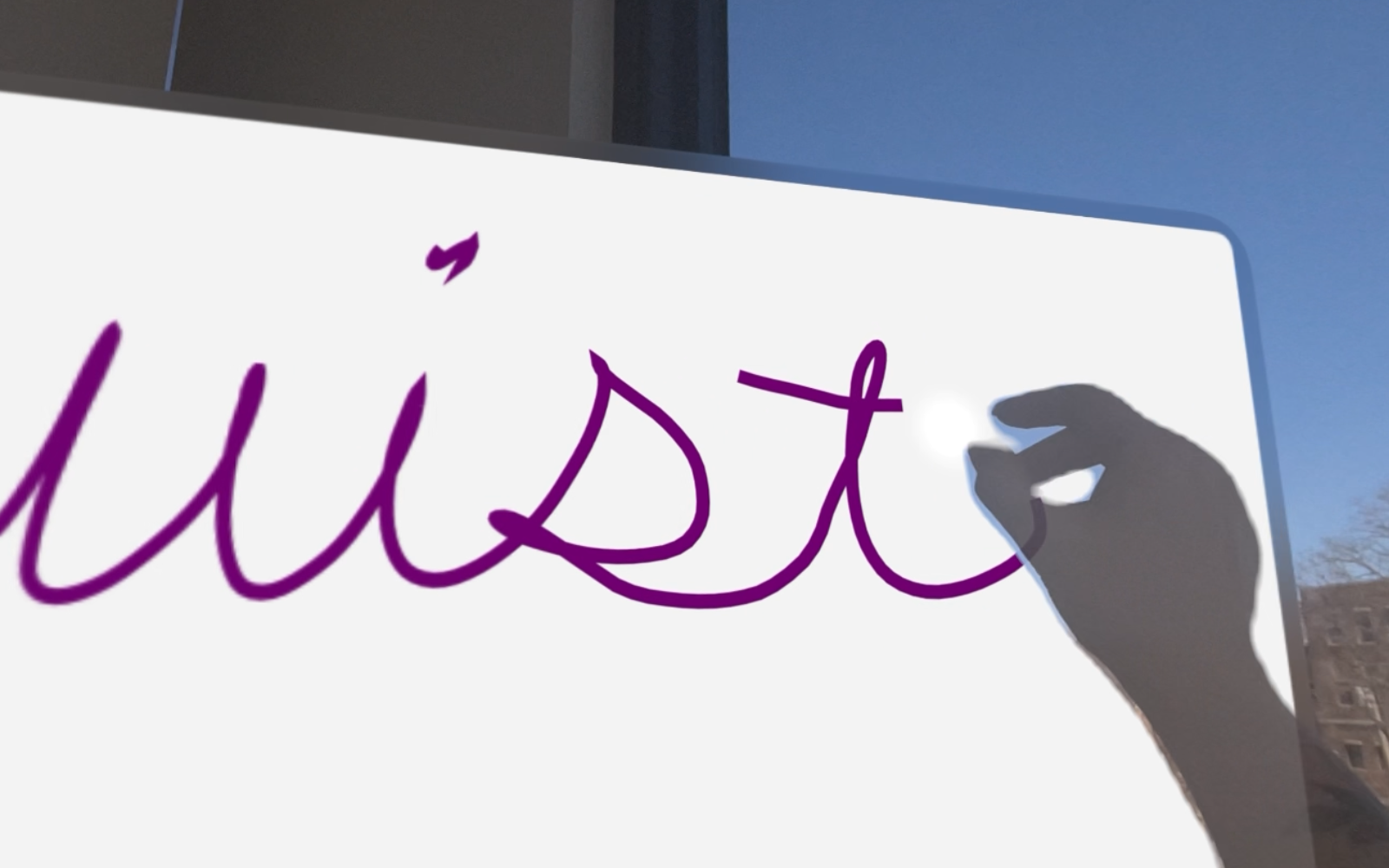}
    \includegraphics[width=0.48\linewidth]{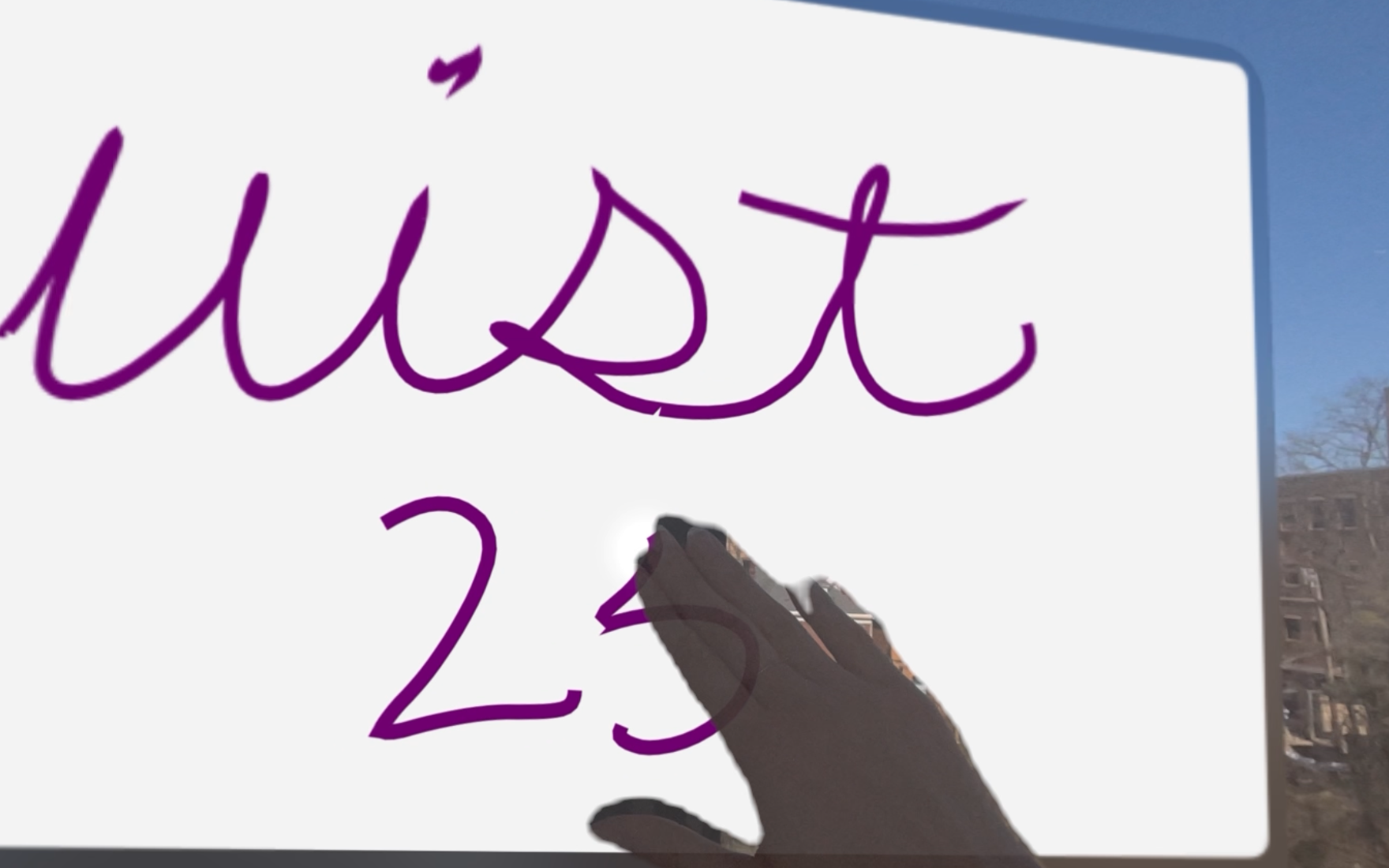}
    \caption{A virtual whiteboard where users draw with a pen-shaped hand (Left) and erase with a wiping gesture (Right).}
    \label{fig:avpdemo}
    \Description{}
\end{figure}

We demonstrate the applicability of our approach to such devices by implementing a creativity app for visionOS on Apple Vision Pro (see \autoref{fig:avpdemo}). In this demo app, the user is able to write on a virtual whiteboard simply by forming the hand into a shape as though writing with a pen. When the user wishes to erase, they can simply change their hand shape and wipe away strokes with the fingers, without the need to select the eraser from a toolbar. We believe this highlights the utility of our method in the context of these growing platforms. The versatility of our approach in recognizing gestures across diverse viewpoints also holds significant potential for this application. Our model was able to successfully use hand keypoints generated by the Apple Vision Pro, despite the fact that this particular device and viewing angle were not present in our training data. This capability unlocks opportunities for a seamless user experience, especially in the growing domain of spatial computing, where hand gestures are a primary mode of interaction.

\section{Discussion}
\label{sec:Discussion}
Here, we expand on the findings of our gesture customization study, discuss limitations, and prospective future directions for this work.

\subsection{Implications for Designing Interactive Gesture Customization Experiences}

Our study findings highlight the importance of incorporating specific interactive features to enhance vision-based gesture customization systems. While our guided gesture registration process simplified sample collection for users, we encountered a challenge where many users inadvertently captured samples that necessitated removal during the review process. This issue is particularly pronounced for dynamic gestures, which rely on precise temporal segmentation. Furthermore, we recognized the value of providing users with feedback on the model's performance during the gesture registration process, helping them decide whether to incorporate a new gesture. Addressing these challenges is pivotal in improving the overall efficiency and accuracy of the customization process. Notably, our approach offers the flexibility for users to augment the training data if they find the model's performance unsatisfactory. As suggested by one user, additional data collection during testing can be guided by the model's confidence in classifying a gesture. When the model lacks confidence in classifying a gesture, the system can prompt the user for confirmation. 
Subsequently, the model can be retrained using this supplementary data, following a methodology akin to active-learning ~\cite{francke2007real}.
Similar techniques have been adopted by Xu et al.~\cite{xu2022enabling} for gestures that are difficult to distinguish or similar to the background class. These techniques encourage users to consider alternative gestures or provide more demonstrations to meet their accuracy needs. This can be handled in the registration process in which alternative gestures can be recommended to the user if the proposed gesture closely resembles existing custom gestures or cannot be distinguished from the background class.

\subsection{Continuous Gesture Recognition}
Determining the start and end of each gesture is crucial for continuous gesture recognition, especially when irrelevant hand movements are intermingled with custom gestures. One approach to address this challenge is to use heuristics that monitor hand acceleration and distinguish between the confidence levels of custom gestures and background class. Specifically, we employ the distance traveled by the hand's centroid on the image to trigger model inference (distance threshold determined empirically). Once the user performs a custom gesture, the confidence score for that gesture increases and decreases, providing a clear segmentation signal. Additionally, irrelevant hand movements are classified as background, enhancing the signal-to-noise ratio for segmentation. Although our paper does not primarily focus on gesture segmentation, these techniques were incorporated into our design tool application and proven effective. Future work could explore additional kinematics-based heuristics, such as hand pose, to further improve segmentation accuracy.

\subsection{Limitations and Future Work}
\subsubsection{Real-World Challenges} Our evaluation studies highlight how our method allows users to create their own set of new gestures. Our method performed well on both a public dataset and data from our own study, demonstrating the promise of combining hand pose estimation, graph transformers, and meta-learning for customizing vision-based gestures. Our method can reject random hand movements (see Section~\ref{sec:BackgroundClass}). However, a key challenge is to differentiate one-handed from two-handed gestures when both hands are in the camera's view. As discussed earlier, we suggest heuristic approaches that track the motion of both hands to address this issue effectively.  Researchers are working on ways to address this problem using gesture segmentation and spotting techniques \cite{wu2016deep,zhu2018continuous,alon2008unified}. This topic area is rich with opportunities to further improve our gesture customization method in the future.

\subsubsection{Generalizability} The assessment involving 21 participants substantiates the effectiveness of our approach. Nevertheless, it is crucial to acknowledge potential biases related to the study population. Notably, our study did not include participants with physical impairments, which limits our ability to evaluate the method's performance in scenarios involving accessibility cases, such as individuals with missing fingers. Future work on gesture customization should be more inclusive to ensure that the technology meets the diverse needs of all users and accommodates a broader range of user capabilities and requirements. This inclusivity will contribute to a more comprehensive and robust evaluation of gesture-based systems and their potential applications.

\subsubsection{Technical Limitations} Our method relies on off-the-shelf hand pose estimation models. This approach has advantages, as it allows issues like hand detection, hand tracking, and feature extraction to be addressed at a higher level. However, errors from these models can propagate through our pipeline and impact the performance of our gesture customization. In the future, we plan to measure and understand these errors better and evaluate pose estimation methods in conjunction with gesture recognition to enhance our system. Finally, although the chosen meta-learning algorithm played a pivotal role in enabling gesture customization, we observed some fluctuations in accuracy across our experiments due to training instability (\textit{e.g.,} training loss fluctuations or not reaching a steady state), as highlighted in previous research~\cite{antoniou2018train}. Additionally, we observed a general decline in accuracy as the number of custom gestures increased. To address these, future investigations could explore alternative meta-learning algorithms, such as metric learning, to enhance the effectiveness of gesture customization and evaluate methods on a higher number of gestures (more than four). 

\subsubsection{Additional Applications}
We believe that our contributions in this paper hold significant potential across various domains. Sign language recognition, for instance, relies on gesture recognition systems, and our customization method empowers users to establish a personalized and robust gesture recognition system, enhancing the accessibility and usability of such technologies.

\section{Conclusion}
In this paper, we introduce a vision-based gesture customization method designed to empower end-users to create personalized gestures with just one demonstration. Our approach initially involves pre-training a graph transformer on a publicly available dataset featuring nine gestures. Subsequently, we present a comprehensive gesture customization method capable of accommodating static, dynamic, one-handed, and two-handed gestures that works across two different viewpoints (\textit{i.e.}, allocentric, egocentric) and has the ability to distinguish gestures from other hand movements. Leveraging meta-learning techniques, we successfully address the challenge of learning from limited samples. In our evaluation, involving 21 participants and a set of 20 newly defined gestures, our method demonstrates an average accuracy of up to 95\% with only one demonstration. We further demonstrate the usability of the real-time implementation of our method through three applications. We believe this work can open doors for users to go beyond pre-existing gestures, empowering them to invent and incorporate new gestures tailored to their unique preferences and capabilities.

\bibliographystyle{ACM-Reference-Format}
\bibliography{sample-base}
\clearpage
\appendix
\section{Complete results}\label{appendix:A}

\begin{figure*}[!hb]
    \centering
\includegraphics[width=0.99\textwidth]{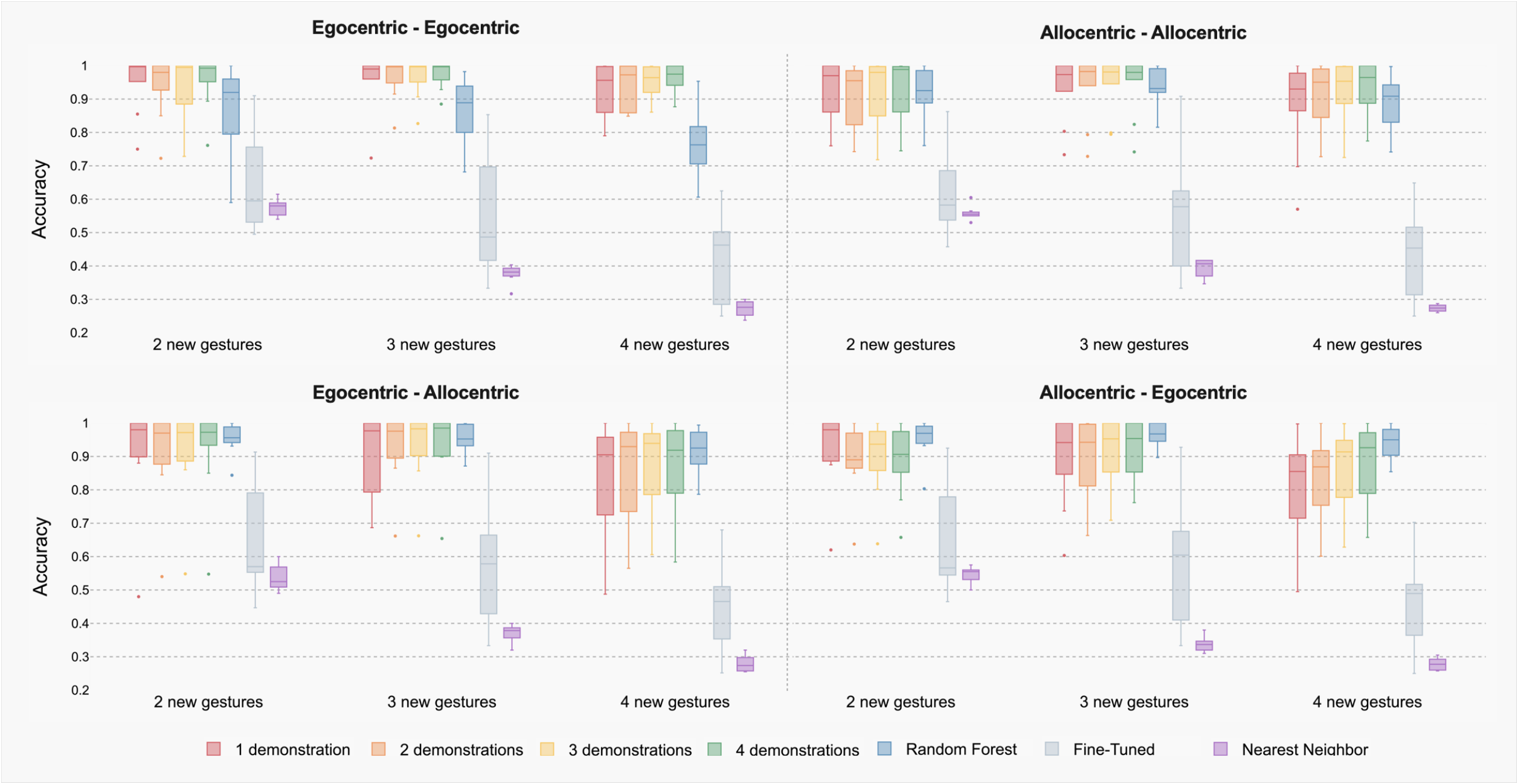}
    \caption{
    Accuracy of our method and baselines without the background class.
    }
    \label{fig:CompleteResults}
    \Description{}
\end{figure*}

\begin{figure*}[!hb]
    \centering
    \includegraphics[width=0.99\textwidth]{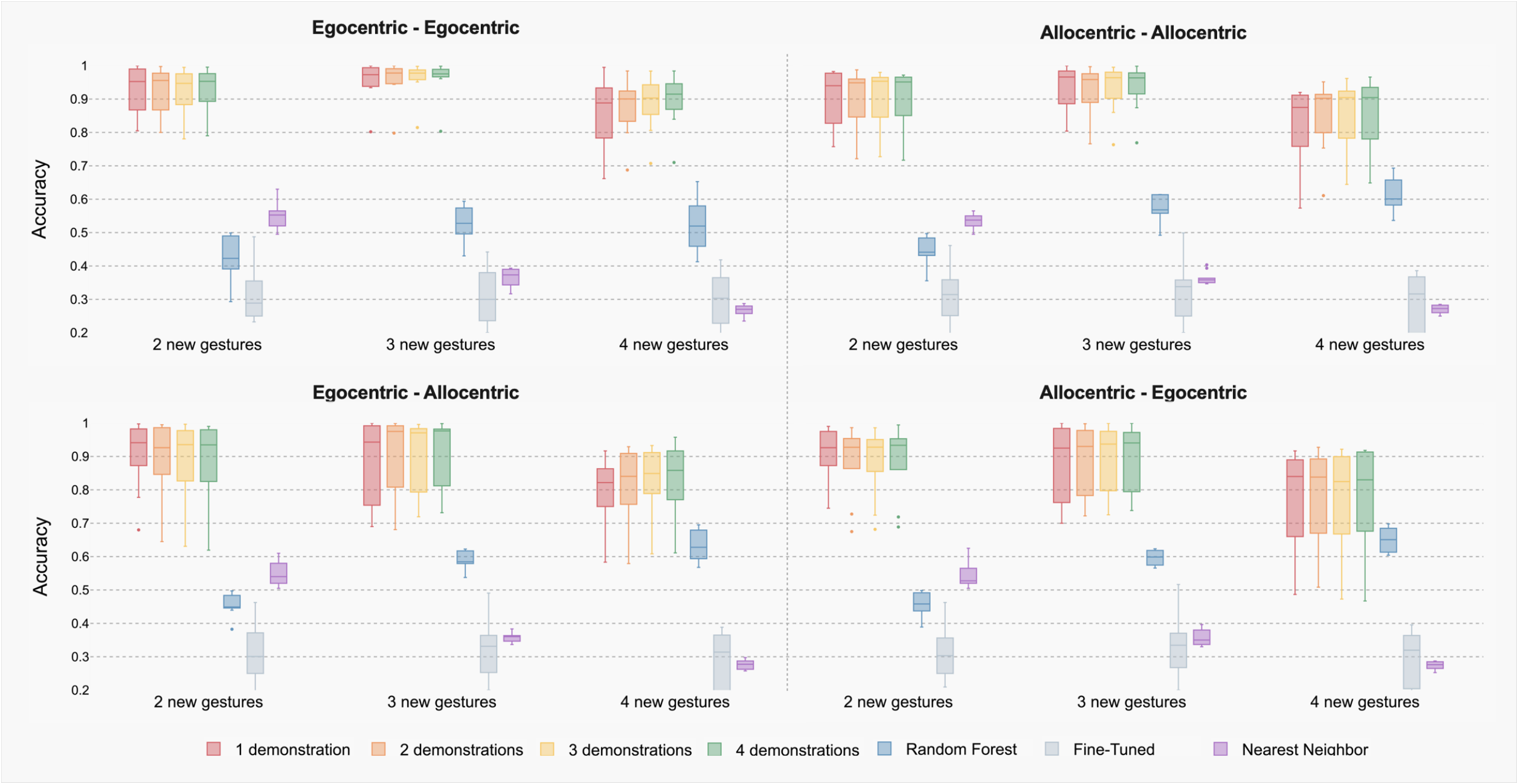}
    \caption{
    Accuracy of our method and baselines with the background class.
    }
    \label{fig:CompleteResultsWithNull}
    \Description{}
\end{figure*}

\end{document}